\documentclass[epj,a4paper]{SVJour}
\usepackage{amssymb,epsf}

\newcommand{\Lp}{{L_{^\parallel}}}
\newcommand{\bu}{{\bf u}}

\newcommand{\br}{{\bf r}}
\newcommand{\ba}{{\bf a}}
\newcommand{\bR}{{\bf R}}
\newcommand{\bq}{{\bf q}}
\newcommand{\bx}{{\bf x}} 
\newcommand{\bk}{{\bf k}}
\newcommand{\qperp}{{\bf q}_{_\perp}}
\newcommand{\DS}{\displaystyle}
\newcommand{\cH}{{\cal H}}
\newcommand{\vdw}{{\rm vdw}}
\newcommand{\eff}{{\it eff}}

\newcommand{\PRB}[1]{Phys.\ Rev.\ B {\bf #1}}

\newcommand{\PRL}[1]{Phys.\ Rev.\ Lett.\ {\bf #1}}

\newcommand{\JETP}[1]{Sov.\ Phys.\ JETP {\bf #1}}

\newcommand{\placefigure}[4]
{
\begin{samepage}
 \begin{figure}[#3]
 \begin{center}
  \leavevmode
  \epsfxsize=#2
  \epsfbox{#1.ps}
 \end{center}  
 \caption{\label{#1} #4}
 \end{figure}
\end{samepage}
}

\begin{document}

\author{Andreas Volmer\inst{1} \and Moshe Schwartz \inst{2}} 
\institute{
  Universit\"at zu K\"oln, Institut f\"ur theoretische Physik,
  Z\"ulpicher Str.\ 77, D-50937 K\"oln, Germany 
  \and
  Tel Aviv University, The Raymond and Beverly Sackler Faculty
  of Exact Sciences,\\ School of Physics and Astronomy, IL-69978 Tel
  Aviv, Israel 
}

\date{\today} 

\title{Van der Waals Interaction between Flux Lines in High-$T_c$
  Superconductors: A Variational Approach} 

\abstract{
  In pure anisotropic or layered superconductors thermal fluctuations
  induce a van der Waals attraction between flux lines.  This
  attraction together with the entropic repulsion has
  interesting consequences for the low field phase diagram; in
  particular, a first order transition from the Meissner phase to the
  mixed state is induced. We introduce a new variational approach that
  allows for the calculation of the effective free energy of the flux
  line lattice on the scale of the mean flux line distance $a$,
  which is based on an expansion of the free energy around the regular
  triangular Abrikosov lattice. Using this technique, the low field
  phase diagram of these materials may be explored. The results of
  this technique are compared with a recent functional RG treatment
  of the same system.
}

\authorrunning{A. Volmer, M. Schwartz}
\titlerunning{VdW Interactions in High-$T_c$ Superconductors: A
  Variational Approach}


\maketitle

\section{Introduction}

The physics of flux lines in high-$T_c$ superconductors at low
magnetic fields is dominated by the competition between the bare
repulsion between flux lines and an entropic repulsion due to thermal
fluctuations. This results in a first order melting transition from
the Abrikosov lattice \cite{Abrikosov} to a liquid phase close to the
lower critical field $H_{c_1}$, as has been predicted some time ago by
Nelson \cite{Nelson}. For mean flux line distances, $a$, larger than
the London penetration depth $\lambda$, the bare interaction decays
exponentially with $a$, while the entropic repulsion decays
algebraically $\sim (\lambda^2/L_T a)^2$. Here, the thermal length
scale $L_T=\Phi_0^2/(16\pi^2T)\approx 2\,$cm$\,$K/$T$ denotes the
length of an isolated flux line segment that shows a thermal mean
square displacement of the order of $\lambda$ \cite{FFH}, and
$\Phi_0=hc/2e$ is the flux quantum carried by each flux line. Close to
the transition, the entropic repulsion dominates over the bare
interaction, leading to a magnetic induction $B$ that vanishes
linearly with the reduced field strength $\tilde
h=(H-H_{c_1})/H_{c_1}$ \cite{Nelson}.

In strongly anisotropic or layered superconductors an additional
interaction between flux lines has been found recently by Blatter and
Geshkenbein (BG) \cite{BlatterGeshkenbein}: Short scale fluctuations
on the scale of pancake vortices lead to an {\em attractive} van der Waals
(vdW) interaction. (This is in fact also the origin of the long range
attraction of a flux line to the surface \cite{Mints}.)  For flux lines
separated by a distance $R$ the strength of this interaction is of the
order $-\lambda^6/(dL_TR^4)$ for $\lambda<R<d/\varepsilon$ and 
$\simeq -\lambda^6/(\varepsilon L_TR^5)$ for
$d/\varepsilon<R<\lambda/\varepsilon$. Here, $\varepsilon^2=m/M\ll 1$
denotes the anisotropy of the material with $m$ and $M$ the effective
masses parallel and perpendicular to the CuO$_2$ plane, respectively,
and $d$ the interlayer spacing. Apart from this thermally induced
attraction, also frozen--in disorder in impure superconductors induces
an attraction between flux lines with the same dependence on the
distance $R$ \cite{NattermannMukherji,bigVDW}. This disorder induced
attraction dominates at very low temperature $T\ll T_{\rm dis}$, where
$T_{\rm dis}$ depends on the disorder strength \cite{FFH}. Here, we
will focus on the opposite case $T\gg T_{\rm dis}$ where it is sufficent
to consider thermal fluctuations.

The competition among the bare, the entropic and the vdW interactions
leads to an interesting phase diagram at low $B$ values. In
particular, the vdW attraction can lead to an instability of the
Abrikosov lattice in the dilute limit $a\gg\lambda$, resulting in a
first order transition between the Meissner and the mixed phase
\cite{BlatterGeshkenbein}.

In order to calculate the low $B$ phase diagram of pure layered
high-$T_c$ superconductors, one has to calculate the Gibbs free energy
density of this system on the scale of the mean flux line distance
$a$,
\begin{equation}
\label{Gibbs}
  g(a;H,T) = f(a,T)-\frac{\varepsilon_\circ \ln\kappa}{a^2} \,\tilde h,
\end{equation}
which has to be minimized with respect to $a$ with the external
magnetic field $H$ fixed. Here, $f(a,T)$ is the free energy density,
$\varepsilon_\circ=(\Phi_0/4\pi\lambda)^2=L_T T/\lambda^2$ is the
basic energy scale, and $\kappa$ is the Ginzburg-Landau parameter.
Note that the bare lower critical field (in the absence of thermal
fluctuations) is given by $H_{c_1}^0=4\pi\varepsilon_\circ
\ln\kappa/\Phi_0$, and that the magnetic flux is related to the flux
line distance via $B\simeq\Phi_0/a^2$. 

It is the aim of the present paper to propose a new variational
approach to obtain this effective Gibbs free energy density, based
on a perturbative expansion in small displacements of the flux
lines around the regular triangular lattice. This expansion is
justified only if we assume that the lattice is stable, i.e., that we
are above the lower melting line below which the flux lines form a
liquid. The variational technique allows for a check
of this assumption by virtue of a Lindemann criterion where the
self-consistently determined flux line fluctuations are compared with
the mean flux line distance $a$. 

In order to motivate our efforts, let us shortly review the approaches
that have been followed so far.  In their first paper
\cite{BlatterGeshkenbein}, BG simply added the bare vdW energy,
evaluated on the scale $a$, to the free energy which contains the bare
and the entropic repulsion mentioned above. In this way, only
contributions from the vdW energy on the length scale $a$ are taken
into account, which leads to a gross underestimation of its influence
because of its rapid decay for $R>\lambda$. In
\cite{NattermannMukherji}, in contrast, only the much more important
contribution on the scale $R_{\rm min}$ was taken into account within
a scaling approach, where $R_{\rm min}$ is the position of the global
minimum of the bare potential $V(R)$. This scaling approach is elegant
and simple and yields the correct qualitative picture, but does not
allow for quantitative predictions.  More recently a functional
renormalization group (RG) calculation has been adapted to the same
problem \cite{bigVDW}; in particular, the low field phase diagram of
Bi$_2$Sr$_2$CaCu$_2$O$_x$ (BiSCCO) was calculated. In that approach,
the bare flux line interaction is renormalized by thermal fluctuations
on all scales between $\lambda$ and $a$, providing an effective free
energy on the scale of $a$. The problem in this latter method is that
the largest scale up to which fluctuations are taken into account is
arbitrary to some extent; we will see below that this flaw is overcome
in a natural way by the technique that we will develop below.

The paper is organized as follows: In Section \ref{sec:method} the
variational method is set up for a arbitrary bare flux line
interactions. The method is applied to a flux line lattice in a
layered high-$T_c$ superconductor in Section \ref{sec:VDW}. There, the
flux line interaction is modeled by the superposition of the bare,
short range repulsion and the long range van der Waals attraction,
using physical parameters typical for BiSCCO.  The data for the free
energy obtained there are used for the minimization of the Gibbs free
energy density in Section \ref{sec:Gibbs}.  We shortly review
the functional RG approach in Section \ref{sec:FRG} and compare the
two methods. Conclusions are drawn in Section \ref{sec:concl}.

\section{Variation of the Free Energy}
\label{sec:method}

Consider a system with $N$ times $M$ flux lines (FLs) in $D=3$
dimensions, confined to a box of base area $L^2$ and height $\Lp$. 
The equilibrium position of line number $(n,m)$ is given by
$\bR_{n,m}=n\,\ba_1+m\,\ba_2$, where $a=|\ba_1|=|\ba_2|$ is the mean
distance between the FLs. Since in equilibrium the FLs
form a triangular lattice we choose  
\begin{equation}
\label{neighbourvectors}
 \ba_n=a \,(\cos(\phi_n)\,{\bf e}_x+\sin(\phi_n)\,{\bf e}_y),\quad
 \phi_n = \frac{\pi}{12}+n\frac{\pi}{3}.
\end{equation}
For $n=1,\ldots,6$ these $\ba_n$ are the distance vectors
$\bR-\bR'$ for all nearest neighbors $\bR'$ of $\bR$ on a
triangular lattice. With this definition of the vectors $\ba_n$ the
equivalence of the $x$ and the $y$ direction is ensured. This results
in the simple form for the matrix $K^{\alpha\beta}(\bq)$ defined
below. 

We write the actual position of line $(n,m)$ at height $z$ as
\begin{equation}
  \br_{n,m}(z) = \bR_{n,m} + \bu_{n,m}(z).
\end{equation} 
The interaction between the vortices $(n,m)$ and $(n',m')$ is
described by the function $V(r(z))$ that depends on the distance at
equal height $z$. We have made the usual assumption that the vortex
coordinates vary slowly with $z$, so that the FL interactions are well
approximated by a potential which is local in $z$; this approximation
is only strictly valid when the vortices are parallel to $z$.

The system is governed by the Hamiltonian
\begin{eqnarray}
  H&=&\frac{\kappa}{2} \int_0^\Lp dz \sum_{n=1}^N\sum_{m=1}^M 
  \left(\partial_z \bu_{n,m}(z)\right)^2 +   \nonumber\\
  && +   \frac12 \int_0^\Lp dz \!\!\!\!\!\!\! \sum_{(n,m)\neq (n',m')} 
  \!\!\!\!\! V(\br_{n,m}(z)-\br_{n',m'}(z)).
\label{Hamiltonian}
\end{eqnarray}
This description of a FL lattice applies in the limit of fluctuations
on scales larger than $\lambda$, with a non-dispersive line stiffness
$\kappa=\varepsilon_l(k_z\ll1/\lambda)$; on
smaller scales, the FL stiffness $\varepsilon_l(k_z)$ in 
anisotropic superconductors is in fact
highly dispersive \cite{blatrev}. The FL interaction $V(R)$ will be
specified in section \ref{sec:VDW}; the following treatment makes no
assumptions about its specific form.

Applying periodic boundary conditions in all 3 spatial directions
(in particular $\bu_{n+N,m}=\bu_{n,m}$ and
$\bu_{n,m+M}=\bu_{n,m}$), Eq.~(\ref{Hamiltonian}) can be rewritten in
Fourier space as  
\begin{eqnarray}
\label{H_Fourier}
  H&=&\frac{\kappa}{2} \sum_{\bq} q_z^2\,
  |\tilde\bu(\bq)|^2 \hfil 
  +\int_0^\Lp \hspace{-.5em} dz \hspace{-1.5em}\sum_{(n,m)\neq (n',m')} 
  \hspace{-1.5em}\int d^2\!k
   \, \times \\
  && \qquad { }\times \,   \tilde V_\bk
  e^{i\bk(\bR_{n,m}-\bR_{n',m'}+\bu_{n,m}(z)-\bu_{n',m'}(z))}. 
  \nonumber \rule{0cm}{3.5ex}
\end{eqnarray}
The wave vector $q_z$ takes the discrete values $0,
2\pi/\Lp,\ldots,\Lambda$. The ultra violet cutoff $\Lambda$ has to be
introduced explicitly in order to avoid a divergence of the free energy.
For the discrete set of $\qperp=(q_x,q_y)$ vectors that are summed
over, we choose    
\begin{equation}
\label{q_nm}
  \qperp =\frac{2}{\sqrt 3}\,\frac{2\pi}{a} 
  \left( 
    \DS\frac{n}{N}\,\cos(\pi/12) - \frac{m}{M}\,\sin(\pi/12)
    \atop
    \DS\frac{m}{M}\,\cos(\pi/12) -
    \frac{n}{N}\,\sin(\pi/12)
    \rule{0cm}{4ex}  
  \right) ,
\end{equation}
with $n=0,\ldots,N-1$ and $m=0,\ldots,M-1$. This is equivalent to
choosing the $\qperp$ vectors from the first Brillouin zone (in the 2
dimensional plane perpendicular to the field direction). The
summation  area in Fourier space, together with the first Brillouin
zone, is illustrated in Fig.~\ref{brilzone}.

The Fourier transform of the fluctuating field $\bu_{n,m}(z)$ is
defined as 
\begin{equation}
\label{u_qomega_disc}
  \tilde\bu(\bq) = \frac1{\sqrt{NM\Lp}} \int_0^\Lp dz \sum_{n,m}
  \bu_{n,m}(z)   e^{-i\qperp\bR_{n,m}-iq_z z},  
\end{equation}
and the Fourier transform of the potential $V(r)$ as $\tilde V_\bk =
\int d^2x\, V(\bx)\, e^{-i\bk\bx}$. 

\placefigure{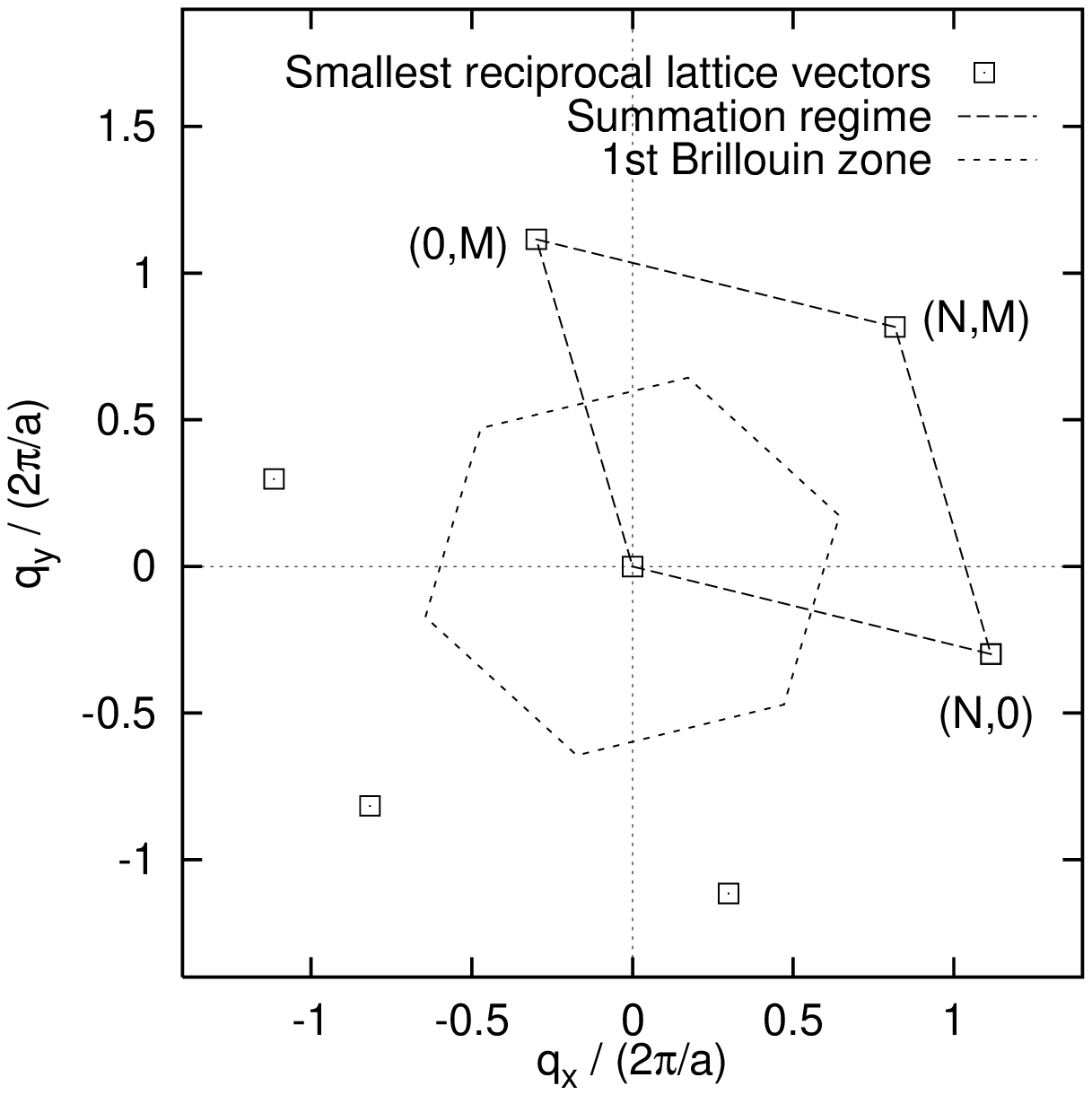}{0.38\textwidth}{bt}{First Brillouin zone and
  summation area in $\qperp$-space. The 
  vectors $\qperp$ with $(n,m)=(N,0)$, $(0,M)$ and $(N,M)$ as defined
  in Eq.~(\protect\ref{q_nm}) are reciprocal lattice vectors.}

Now, we define a Gaussian variational Hamiltonian
\begin{equation}
  H_0 = \frac12 \sum_{\bq} K^{\alpha\beta}(\bq)
        \tilde u_{\alpha}(\bq) \tilde u_\beta(-\bq),
\end{equation}
with summation over the Cartesian components $\alpha,\beta=1,2$, and
rewrite the original Hamiltonian as 
\begin{equation}
\label{H_division}
  H = H_0 + (H-H_0).
\end{equation}
Let $\langle \ldots \rangle_0$ denote thermal averages with respect to
$H_0$. Clearly, 
\begin{equation}
\label{uu_corr}
  \langle u_\alpha(\bq)u_\beta(\bq') \rangle_0 \,
  =  k_B T \;\delta^3_{\bq+\bq'}   \,
  \tilde K^{\alpha\beta}(\bq),
\end{equation}
where $\tilde K(\bq)$ is the Matrix inverse of $K(\bq)$, i.e.,\\ 
\mbox{$K^{\alpha\beta}(\bq)\tilde K^{\beta\gamma}(\bq)=\delta_{\alpha\gamma}$}.

The correlation function $K(\bq)$ is a 2x2 matrix:
\begin{equation}
  K(\bq) = \left(
    \begin{array}{cc}
      K_D(\bq) & K_O(\bq) \\
      K_O^*(\bq) & K_D(\bq) \rule{0cm}{3ex}
    \end{array}
  \right).
\end{equation}
While the diagonal terms $K_D(\bq)$ are real, the off-diagonal
part $K_O(\bq)$ may in principle take on complex
values. In a straightforward but tedious calculation it can be shown
however that $K_O(\bq)$ is actually 
real, too. For completeness, we give the explicit expression for the
matrix inverse $\tilde K(\bq)$: 
\begin{equation}
  \tilde K_{D(O)}(\bq) = +(-)\;\frac{K_{D(O)}(\bq)}
  {K_D^2(\bq)-K_O^2(\bq)} .
\end{equation}

As stated in the introduction, our aim is to minimize the Gibbs free
energy density $g(a;H,T)$ as defined in Eq.~(\ref{Gibbs}) as a
function of the mean FL separation $a$. Let us first consider the free
energy $F=F(a,T)$. We define an approximate free energy $F^*$ that is
obtained by a first order perturbation expansion around $F_0$, the
free energy associated with $H_0$,
\begin{equation}
  F^* = F_0 + \langle H-H_0 \rangle_0.
\end{equation}
It is well known \cite{Chaikin} that for any choice of $H_0$, $F^*$ is
an upper bound on the true free energy $F$. Therefore minimization of
$F^*$ with respect to the parameters defining $H_0$, $K_D(\bq)$ and
$K_O(\bq)$, yields the best approximation of its kind, namely $H_0$
corresponding to the optimal parameters is the best fluctuating
lattice Hamiltonian that may be used to describe our physical system. 
The free energy $F_0$ is given by
\begin{equation}
\label{F0}
  F_0 = -\beta^{-1}\ln\int {\cal D}[\bu] \, e^{-\beta H_0}
   = \frac1{2\beta} \sum_{\bq}
  \ln\mbox{det}(\beta K(\bq)) .
\end{equation}

Next, the expectation value $\langle H-H_0 \rangle_0$ shall be
calculated.  To proceed, it is useful to rewrite the last term in the
interaction part of (\ref{H_Fourier}) using the Fourier representation
(\ref{u_qomega_disc}). 
Then, the averaging can be immediately carried out, leading to
\begin{eqnarray}
\label{H_H0}
  \langle H-H_0 \rangle_0 &=&
  \frac{\kappa}{2}\sum_{\bq} q_z^2 \;
  {\rm Tr}\,[\,\tilde K(\bq)\,] - \;\Lp NM\frac{\Lambda}{2\pi} k_BT\;
  +\nonumber\\  
  &&\hspace{-3em}
  + \; \frac{\Lp NM}{2}  
     \sum_{\ba} \int_\bk \,\tilde V_\bk \,e^{i\bk\ba} \times \\
  && \exp\left(-\frac{2 k_\alpha k_\beta }{NM\Lp} 
       \sum_{\bq} 
       \sin^2(\qperp\!\! \cdot\! \ba/2) \;
       \tilde K^{\alpha\beta}(\bq)
     \right) .
    \nonumber 
\end{eqnarray}
We restrict the sum $\sum_{\ba}$ to a summation over nearest
neighbors, which is a good approximation in the dilute limit we are
interested in; in principle, one can also include next-nearest
neighbors, etc. For the triangular lattice, it is convenient to
choose the 6 vectors $\ba_n$ from Eq.~(\ref{neighbourvectors}), with
$n=1,\ldots,6$.

In the following, we choose the temperature $k_BT=\beta^{-1}$ as the
basic energy scale, and introduce the parameters
$\tilde\kappa=\beta\kappa$, the reduced potential $v(\br)=\beta V(\br)$
and the reduced Hamiltonian $\cH=\beta H$. Finally, we replace
$\beta K(\bq)$ by $K(\bq)$.

The variational parameters $K_D(\bq)$ and $K_O(\bq)$ are
determined by minimizing $F^*$:
\begin{equation}
  \frac{\partial F^*}{\partial
  K_{D(O)}(\bq)}  \;\stackrel{\textstyle !}{=}\, 0 .
\label{K_selfcons}
\end{equation}
This leads to
\begin{eqnarray}
\label{K_qomega}
  K_D(\bq) &=& \tilde\kappa q_z^2 \,-\, 2\sum_\ba
  \sin^2(\qperp\cdot\ba/2)\,A_\ba \\ 
  K_O(\bq) &=& - 2\sum_\ba \sin^2(\qperp\cdot\ba/2)\,B_\ba 
\end{eqnarray}
where we have defined
\begin{eqnarray}
\label{AB_def}
  A_{\ba} &=& \frac12\int_\bk \bk^2 \, \tilde v(\bk)\, e^{-i\bk\ba}
  e^{-\frac12 k_\alpha   M^{\alpha\beta}_\ba k_\beta}
  \qquad \mbox{and}\\
  B_{\ba} &=& \frac12\int_\bk 2k_xk_y \, \tilde v(\bk)\, e^{-i\bk\ba}
  e^{-\frac12 k_\alpha   M^{\alpha\beta}_\ba k_\beta}. 
\end{eqnarray}
By comparison with Eq.~(\ref{H_H0}), the 2x2 matrix $M_\ba$ is defined as
\begin{equation}
\label{M_a_def}
  M_{\ba}^{\alpha\beta}
\label{M_fluctuations}
  = \frac{4}{MN\Lp} \sum_{\bq}
  \sin^2(\qperp\cdot\ba/2)\,\langle u_\alpha(\bq)
  u_\beta(-\bq) \rangle_0,
\end{equation}
where we have plugged in the representation (\ref{uu_corr}) of the
matrix inverse $\tilde K(\bq)$.  Again, $M_{\ba,D}$ and $M_{\ba,O}$
are the diagonal and off-diagonal terms, respectively. $\tilde M_\ba$
denotes the matrix inverse of $M_\ba$.  Eqs.~(\ref{K_selfcons}) to
(\ref{M_a_def}) serve as a self-consistent set of equations for
$K(\bq)$.

\section{Van der Waals Interaction}
\label{sec:VDW}

To be specific, we take the bare potential $v(R)$ to be given by the
superposition of the short range attractive and the long range
repulsive interaction describing the direct vortex-vortex interaction in
the extremely decoupled limit $\varepsilon\rightarrow 0$,
\begin{equation}
\label{bareV0spec}
  v(R) = v_0 \left( K_0(R/\lambda) - a_{\rm vdw} \phi(R/\lambda)
    \frac{\lambda^4}{R^4} \right),
\end{equation}
where $v_0=2\varepsilon_\circ/k_B T$ measures the amplitude of the direct
interaction between flux lines, and $a_{\rm vdw}$ determines the
strength of the thermal vdW attraction.  $\phi(x)$ is a function that
smoothly cuts off the power law part for $R\lesssim\lambda$, which we
have defined as
\begin{equation}
  \phi(x) = \left\{
    \begin{array}{ll}
      0, & x \le x_1 \\
      \frac14
        \left[ 
          1+\sin\left(\pi\frac{x-(x_1+x_2)/2}{x_2-x_1}\right)
        \right]^2, 
        & x_1<x<x_2 \\   
      1, & x \ge x_2
    \end{array}
  \right.
\label{CutOff}
\end{equation}
with $x_1=1$ and $x_2=5$. The choice of the cutoff function, as well as
the actual values of $x_1$ and $x_2$, is to some extent
arbitrary; $x_1=1$ is however an obvious choice, and $x_2$ has to be
chosen such that the cutoff is not too sharp and,
on the other hand, does not influence the form of the potential in
the vicinity of the minimum for those values of $a_{\rm vdw}$ that are
physically meaningful \cite{bigVDW}.

In order to make quantitative predictions for the low-field phase
diagram of layered high-$T_c$ superconductors, we identify the
parameters introduced above with physical parameters characterizing
those systems: The elastic constant is
$\tilde\kappa=\varepsilon_\circ/2k_BT$ in the long wavelength regime
$\lambda k_z\ll 1$ \cite{blatrev}, and the amplitude of the
vdW attraction (relative to $v_0$) is given by $a_\vdw \approx
k_BT/(2\varepsilon_\circ d\ln^2(\pi\lambda/d))$, where $d$ is the layer
spacing \cite{BlatterGeshkenbein,bigVDW}.

Let us quantify these parameters for a specific highly anisotropic
material, BiSCCO. This 
superconductor is characterized by the London penetration depth
$\lambda\approx 2000\,${\AA}, a Ginzburg-Landau parameter
$\kappa\approx 100$, a layer spacing $d\approx 15\,${\AA} and
an anisotropy parameter $\varepsilon\approx 1/300$
\cite{blatrev}. Together with the thermal length $L_T\approx 2\,{\rm
  cm}\,{\rm K}/T$, we find for our model parameters the values
$\lambda\tilde\kappa\approx 10^5\,$K$/T$, $v_0=4\tilde\kappa$, and
$a_\vdw\approx 2\times 10^{-5}\,T/$K. At $T=100\,$K, which is of the
order of the critical temperature $T_c$, we finally have
$\lambda\tilde\kappa\approx 10^3$ and $a_\vdw\approx 2\times
10^{-3}$. We want to feed the potential (\ref{bareV0spec}) with
$\tilde\kappa$ and $v_0$ discussed above into the variational
procedure. The London penetration depth $\lambda$ is
chosen to set the length scale in the direction perpendicular to the
FLs, i.e., all lengths are measured in terms of $\lambda$.

Although the vdW amplitude $a_{\vdw}$ is determined by material
parameters and the temperature as noted above, we will tune this 
parameter here in order to find whether a 'critical' value
$a^*_{\vdw}$ exists where the phase transition from the Meissner phase
to the mixed phase changes its character from a second order (for
$a<a^*_{\vdw}$) to a first order transition (for
$a>a^*_{\vdw}$).

With these parameter values and the potential (\ref{bareV0spec}), we
have solved the set of self-consistent equations (\ref{K_selfcons}) to
(\ref{M_a_def}) numerically for different values of the mean spacing
$a$ with fixed FL number $N\times M$ and fixed vertical length $\Lp$.
In order to speed up computation, the sums over $q_z$ were replaced by
integrals.

First, we present the fluctuation matrices (\ref{M_fluctuations}) as a
function of the mean FL distance $a$.  There are 6 fluctuation
matrices corresponding to the 6 nearest neighbor vectors given by
Eq.~(\ref{neighbourvectors}). The special choice of these vectors
$\ba_n$, however, yields only two different matrices, namely
\begin{eqnarray}
\label{Mi_defs}
  M^{(1)}&\equiv& M_{\ba_n} \qquad \mbox{for~} n=1,3,4,6 \mbox{~and}
  \nonumber \\
  M^{(2)} &\equiv& M_{\ba_{n}}\qquad \mbox{for~} n=2,5.
\end{eqnarray}
With $M^{(1)}$ and $M^{(2)}$ each having two
independent entries (the diagonal and off-diagonal terms $M_D$ and
$M_O$, respectively), we have to solve for a total of 4 variables.
The implementation made use of a multi-dimensional Newton-Raphson
method \cite{NumRecipes}. Numerical solutions for these four
observables are shown as a function of the mean spacing $a$ in
Fig.~\ref{fluctVDW}. 
The data shown in this figure stems from
calculations with the vdW amplitude $a_\vdw=10^{-6}$;
the corresponding data for higher values of $a_\vdw$ is not
significantly different.

For values of $a$ much smaller than the minimum position of the bare
potential (which is at $R_{\rm min}\approx 27\lambda$ for the vdW
amplitude chosen here) fluctuations are strongly suppressed by the
direct repulsion from the nearest neighbors. The fluctuations grow
rapidly around $a\approx 25\lambda$, because in that regime
the strong repulsion is neutralized by the vdW attraction. For a large
intermediate range, the fluctuations scale almost $\sim a^2$.  Note that
in that plot, this quadratic scaling would correspond to a
horizontal line.  Actually, the slope is slightly negative, because
the attraction continuously decreases for larger distances. This
tendency is most obvious in the plot of the fluctuations 
\begin{equation}
  \langle \bu(0)^2\rangle = \frac{2}{MN\Lp}
  \sum_{\bq} \tilde K_D(\bq)
\end{equation}
as a function of $a$ in Fig.~\ref{fluctU2VDW}.  For $a\gtrsim
150\lambda$, the size of the fluctuations saturates. This is due to
the finite system size $\Lp$ in the $z$-direction; in this limit, the
contribution of the potential energy to the free energy density is
negligible, leaving only the kinetic term. This leads to
\begin{eqnarray}
\label{M_D_upperbound}
  M_{\ba,D} &=& \langle (u_x(\bx+\ba,t)-u_x(\bx,t))^2 \rangle
  \nonumber \\
  &\simeq& 2 \int_{2\pi/\Lp}^{\Lambda}
  \frac{d q_z}{2\pi} \frac{1}{\tilde\kappa q_z^2} 
  \approx \frac{1}{2\pi^2} \frac{\Lp}{\tilde\kappa},
\end{eqnarray}
where we have assumed that the correlations between neighbors are
small and that $\Lp\Lambda/2\pi\gg 1$. Finite size effects will hence
occur when the mean vortex distance $a$ becomes of the order of
$\sqrt{\Lp/2\pi^2\tilde\kappa}$.  In this regime, the off-diagonal
terms $M_O^{(1)}$ and $M_O^{(2)}$ vanish due to the effective
rotational invariance for each FL in this limit, where
the FLs no longer interact with each other.

\placefigure{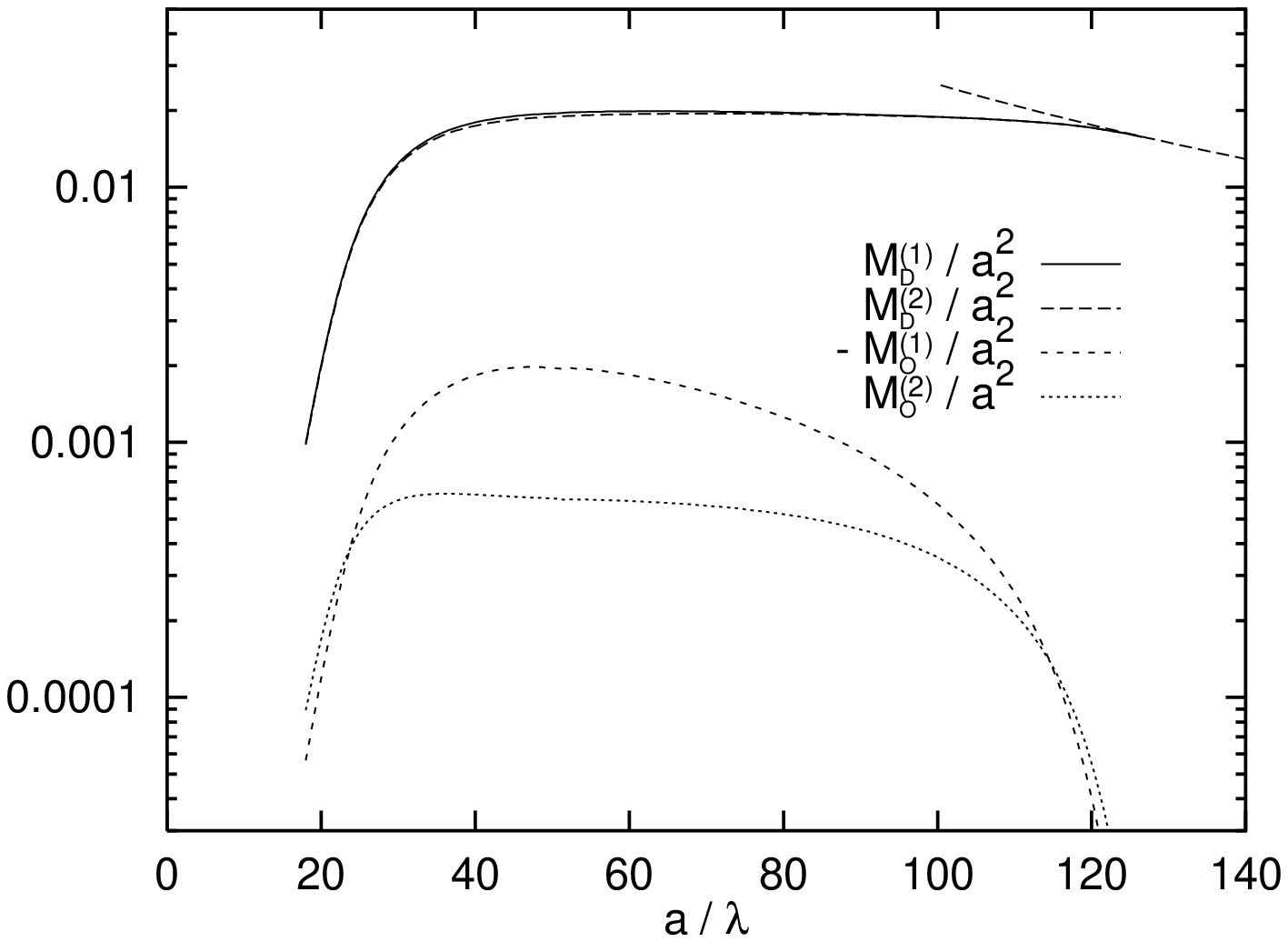}{0.49\textwidth}{tb}{
  Fluctuations $M^{(i)}$ divided by $a^2$, as a function of
  the spacing $a$. The two curves $M^{(1)}_D$ and $M^{(2)}_D$ are hardly
  distinguishable. The vdW amplitude has been set to $a_\vdw=10^{-6}$,
  and the other model parameters to $\tilde\kappa=10^3$, $v_0=4\times 10^3$,
  $N=M=100$, $\Lp=5\times 10^6$, $\Lambda/\pi=400$. In the upper right
  corner of the figure, the function $\Lp/2\pi^2\tilde\kappa a^2$ has been
  added ({\em dashed line}) which gives the upper bound for
  $M^{(i)}_D$, as derived in Eq.\ (\protect\ref{M_D_upperbound}).}
\placefigure{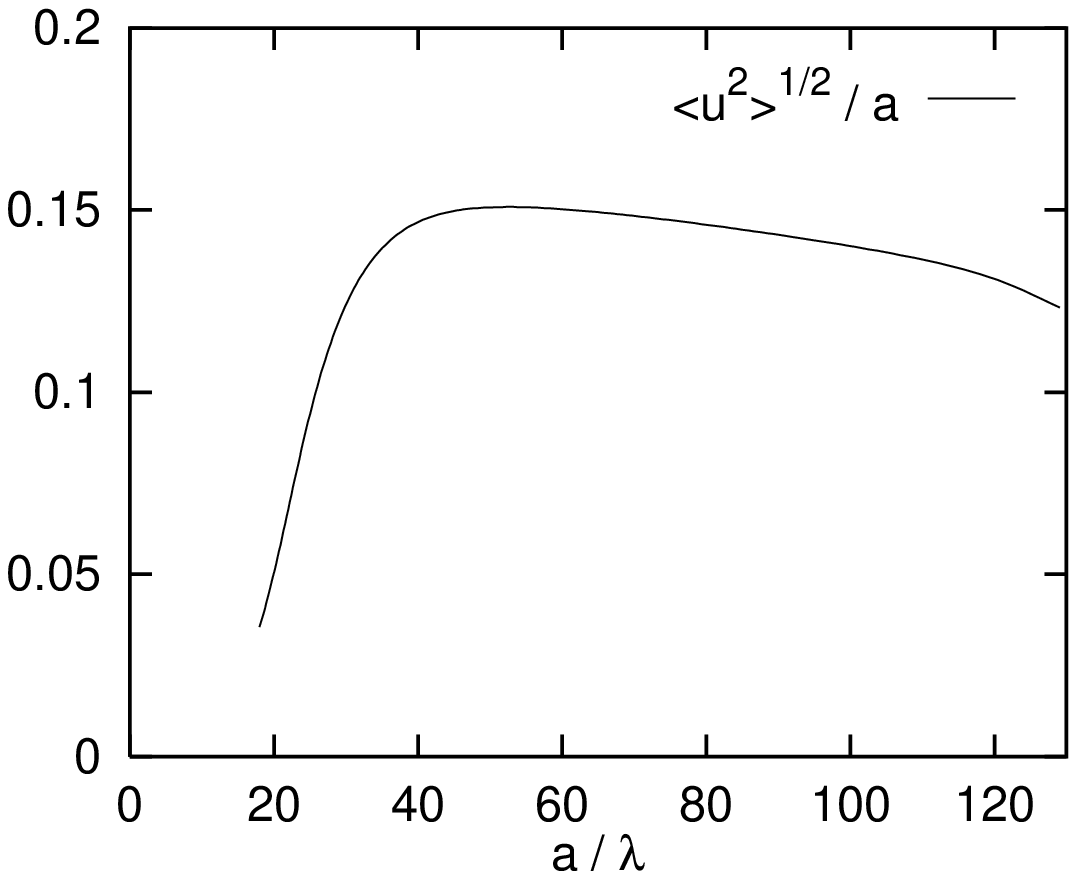}{0.35\textwidth}{tb}{The fluctuations
  $(\langle \bu^2(0)\rangle)^{1/2}/a$,   as a function of $a$, from
  the same numerical data as in Fig.~\protect\ref{fluctVDW}.}

\vspace{1ex}

\section[Minimization of the Gibbs Free Energy]{Minimization of the
  Gibbs Free\\ Energy}
\label{sec:Gibbs}

We will now use the effective free energy density $f(a,T) = F(a,T)/\Lp
NMa^2$, where the free energy $F(a,T)$ is taken to be the minimal value of
$F^*$, to calculate the Gibbs free energy density
\begin{equation}
\label{GibbsDensity}
  g(a;\mu,T) = f(a,T) + \mu/a^2
\end{equation}
which has to be minimized with respect to the mean FL distance
$a$. For convenience, we have introduced 
the 'chemical potential' $\mu\equiv -\varepsilon_\circ\tilde
h\ln\kappa$, where $\tilde h = (H-H_{c_1})/H_{c_1}$ is the deviation
of the applied magnetic field from the lower critical field.

In Figs.\ \ref{trans2VDW} and \ref{trans1VDW}, the Gibbs free energy
density $g(a;\mu,T)$ as defined in (\ref{GibbsDensity}) is plotted for
two different values of $a_\vdw$.  For these two values of the vdW
amplitude, the transition from the Meissner state to the mixed phase
has different characteristics: For $a_\vdw=10^{-5}$, the transition is
continuous because the minimum position of the potential is
continuously shifted to larger values with growing $\mu$,
corresponding to a second order phase transition.

For $a_\vdw=10^{-4}$, on the other hand, the transition is
discontinuous and hence first order, because for $\mu$ in the vicinity
of $\mu_c$, two minima emerge: one at a finite value of $a$ and the
other at $a=\infty$.  The magnetization $B\sim 1/a_{\rm min}^2$ as a
function of $H$ thus shows a first order transition from the Meissner
phase to the mixed phase at a finite magnetization $B_v$.  Hence, the
principal result from the study by BG 
\cite{BlatterGeshkenbein} and from the functional RG treatment
\cite{bigVDW} is reproduced by the variational approach.

\placefigure{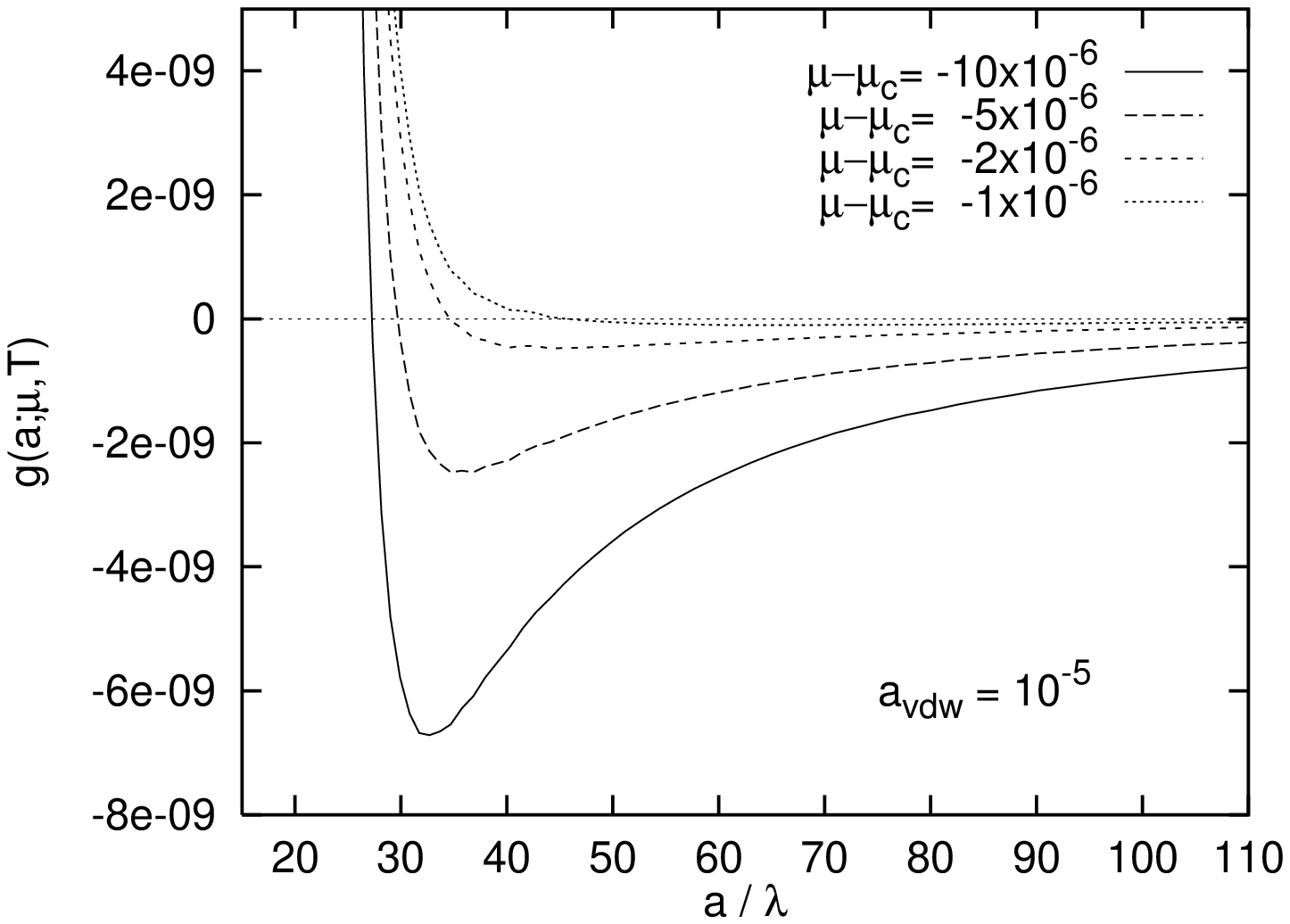}{0.49\textwidth}{tb}
{ Gibbs free energy density $g(a;\mu,T)$ for several values of $\mu$,
  with $a_\vdw=10^{-5}$. 
  The data reflects a second order transition from the Meissner 
  phase to the mixed phase. 
}
\placefigure{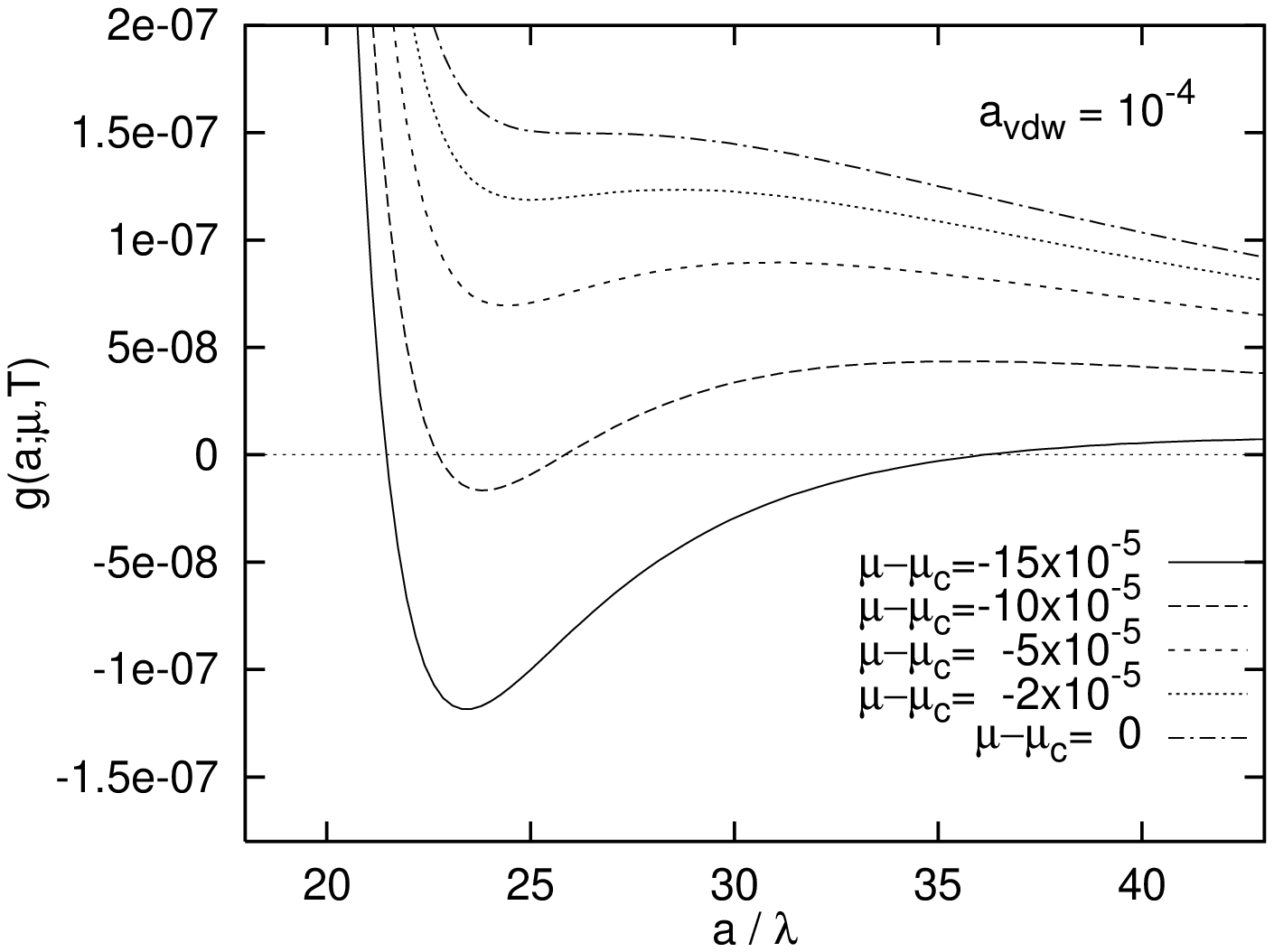}{0.49\textwidth}{tb}
{ Gibbs free energy density $g(a;\mu,T)$, now with
  $a_\vdw=10^{-4}$. The figure reflects a first 
  order transition  from the Meissner phase to the mixed phase. 
}

The transition is illustrated in Fig.\ \ref{BversusMuVDW}, where
$1/a_{\rm min}^2(\mu)$ is plotted as a function of $\mu-\mu_c$. For
each value of $a_\vdw$, $\mu_c$ has been determined individually such
that $g(a;\mu>\mu_c,T)$ has {\em no} minimum at {\em finite} $a$,
while it {\em does} have such a minimum for $\mu<\mu_c$.  For
$a_\vdw<a^*_\vdw$, the magnetic induction is proportional to
$\mu-\mu_c$. This scaling arises from the competition between the bare
repulsive interaction and the entropic repulsion alone.

For the parameters chosen above, the 'critical' value of $a_\vdw$
separating the two regimes is found to be $a^*_\vdw \approx 2\times
10^{-5}$. As stated above, the physical value of the amplitude of the
thermally induced van der Waals attraction is of the order $a_\vdw\approx
10^{-3}$, hence deep in the regime where the transition -- according
to the variational procedure -- is first order.  From the data in
Fig.\ \ref{BversusMuVDW}, we read off that the mean separation between
FLs just above the transition between the Meissner phase and
the mixed phase is $a_{\rm min} \approx 20\lambda$. For BiSCCO, this
corresponds to a magnetic induction $B_v\approx
500\,$G$/(a_{\rm min}/\lambda)^2 \approx 1.2\,$G \cite{blatrev}.

\placefigure{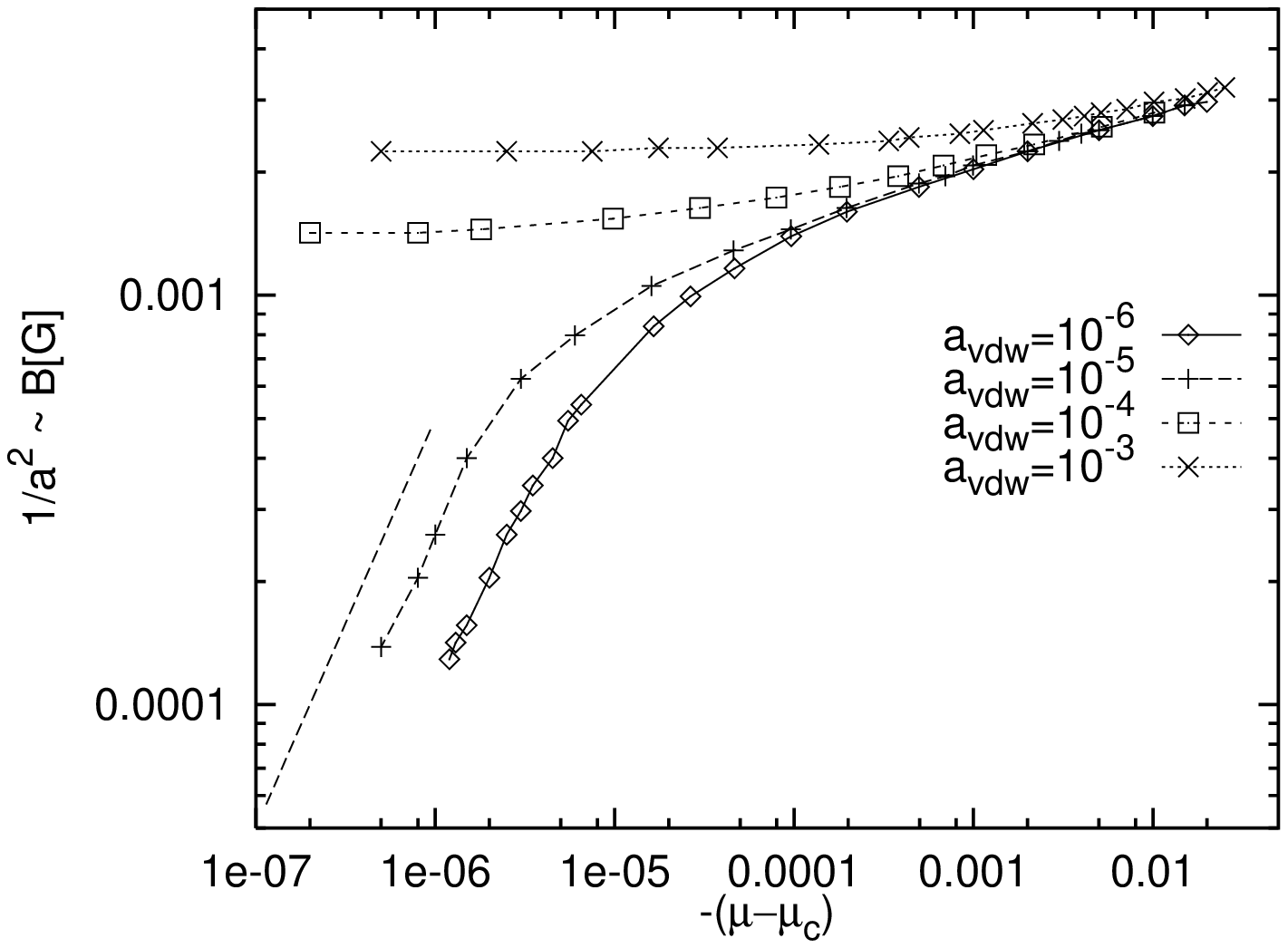}{0.49\textwidth}{tb}
{ $1/a_{\rm min}^2$ versus $\mu-\mu_c$ for several values of
  $a_\vdw$. In the lower left corner, a ({\em dashed}) line has been
  added that shows a linear scaling of the magnetic induction
  $B$ as a function of $\mu-\mu_c$, revealing the linear scaling of
  the two lower curves which correspond to the two lowest values of
  the vdW amplitude. The other curves approach a finite value $1/a^{*2}$
  for   $\mu\rightarrow \mu_c$, thus determining the van der Waals magnetic
  induction $B_v=\Phi_0/a^{*2}$.}

Now, let us face the question whether the basic assumption of the
variational approach, i.e., the existence of a regular triangular
lattice, is valid for these parameters. This can be checked using
the self-consistently determined fluctuations by applying a Lindemann
criterion \cite{Lindemann,NelsonSeung}. In particular, the elastic
structure melts when the displacement between two neighbors becomes 
\begin{equation}
  \langle (\bu_n(z)-\bu_m(z))^2 \rangle \ge c_L^2 a^2,
\end{equation}
where $n$ and $m$ denote two neighboring vortices, $a$ is the lattice
constant, and $c_L$ is the heuristically determined
Lindemann number. In a recent numerical study on FL lattice melting
\cite{Nordborg}, this number has been found to be $c_L\approx 0.25$,
in agreement with usual estimates for vortex lattice melting
\cite{blatrev}.  With the values represented in Fig.\ \ref{fluctU2VDW},
we are hence above the lower melting line where the lattice is
still stable \cite{Note}.

In principle, the variational technique can now be used to explore the
whole low field phase diagram of layered superconductors, tuning the
temperature $T$. We will restrict ourselves here however to the data
shown above, since we do not expect results for the phase diagram that
differ qualitatively from those obtained by the functional
renormalization group treatment to be discussed below. Instead, we
will compare the two techniques and discuss their relative strengths
and flaws.

\section{Comparison with the functional RG}
\label{sec:FRG}

As noted in the introduction, the problem considered here has
recently been addressed by a functional renormalization group
calculation \cite{bigVDW}. In that approach, the FL interaction
potential, Eq.~(\ref{bareV0spec}), is renormalized by short scale
thermal fluctuations, analogous to the first RG treatment of the
purely repulsive short range FL interaction by Nelson and Seung
\cite{NelsonSeung}. The starting point of this technique is the bare
Hamiltonian (\ref{Hamiltonian}) with two interacting FLs. Due
to the presence of neighbor vortices in a typical distance $a$,
transverse fluctuations are assumed to be confined to this length
scale. Hence, thermal fluctuations are integrated out by means of a
momentum shell RG up to this scale, $a$, applying (nonlinear)
recursion relations that are closely related to those that were
established in the context of the wetting transition \cite{Lipowsky}.
(Note that this argument applies to the system in a liquid state; in
the solid state, the fluctuations are confined to a shorter scale $<
c_L a$, as we have seen above.)  This procedure renormalizes the
interaction potential, leading to an effective potential $V^{\rm
  eff}(R)$ on the scale $a$.

For large $a\gg\lambda$, the RG takes us into a region of weak
coupling and high vortex densities \cite{NelsonSeung}, where 
the effective Gibbs free energy density simply reads
$g(a;\mu,T)=(\frac{z}{2}V^{\rm eff}(a)+\mu)/a^2$. $z=6$ is the
number of nearest neighbors. Again, $g(a;\mu,T)$ has to be minimized
with respect to $a$ in order to obtain the Gibbs free energy density
as a function of $\mu$ and $T$, hence on this level the analysis of
the phase diagram is identical to that employed within the variational
technique. This enables us to directly compare the results of the two
approaches. 

Returning to the result for $B_v$ obtained above for BiSCCO at
$T\approx 100\,$K, one finds that this value is considerably higher than
$B_v^{\rm RG}\approx 0.2\,$G as calculated from the functional
RG data \cite{bigVDW}. This discrepancy is consistent with the fact
that the value for $a^*_\vdw$ determined above is much smaller than
the value $a^*_\vdw\approx 2\times 10^{-3}$ which results from the RG
calculations.

The reason for the discrepancy between the two approaches is given by
the fact that they include fluctuations up to different maximal
scales, as mentioned above. In particular, we have found here that
fluctuations are confined to $\sqrt{\langle \bu^2 \rangle} \approx
(0.10\ldots 0.15)a$.  Hence, the FLs have a much smaller
contact probability than assumed in the RG procedure, and the strong
short range repulsion does consequently contribute less to the
effective interaction energy, rendering the effective potential more
attractive.

This statement can be made more quantitative by ad\-apting the data 
from the functional RG to the situation where the FLs form a lattice by
confining the integration to 
scales $\le\frac{1}{\eta}\,a$ with a constant $\eta>1$. 
An example for the effective interaction $V^{\eff}_{\eta}$ obtained in
this way is shown in Fig.~\ref{veta_frg4}. 

\placefigure{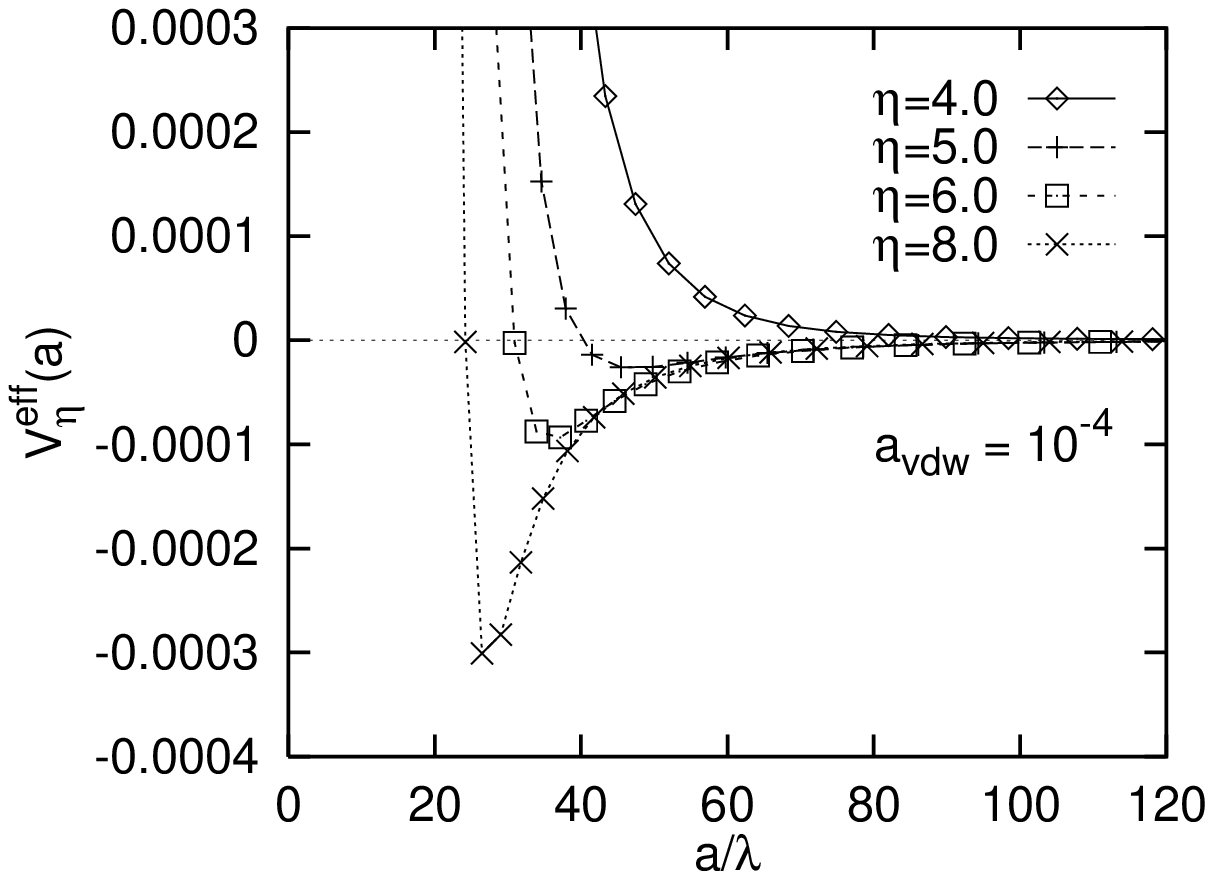}{0.49\textwidth}{tb}{
The effective interaction energy $V^{\eff}_{\eta}(a)$ which results
from the functional RG procedure when stopping the integration on the
scale $a/\eta$. The data result from numerical functional RG
calculations with a rescaling factor $b=1.2$ and $v_0=2\times 10^6$,
corresponding to the temperature $T=100\,$K. The vdW
amplitude is set to the values $a_{\vdw}=10^{-4}$.}

Indeed, the figure reveals that $V^\eta_{\eff}(a)$ exhibits a global
minimum at a finite length scale $a$ for large enough $\eta$,
while it lacks such a minimum for smaller $\eta\approx 1$. With
growing $\eta$, the position of the minimum decreases, corresponding to
larger values $B_v$.

Let us compare the results which are shown in Figs.\ \ref{trans1VDW}
and \ref{veta_frg4} from the variational and the RG approach,
respectively, corresponding to the same temperature $T=100\,$K and the
same vdW amplitude $a_{vdw}=10^{-4}$.  For this comparison, we
determine the appropriate coefficient $\eta$ from the magnitude
$\langle \bu(0)^2 \rangle$ of the fluctuations, as measured in the
variational calculation, with $a$ set to the length $a_{\min}$ that
corresponds to the position of the free energy minimum. For the
present set of parameters, $a_{\min}\approx 25\lambda$ (see Fig.\ 
\ref{trans1VDW}). The corresponding fluctuations have been determined
to be $\langle \bu(0)^2 \rangle^{1/2} \approx (0.1\pm 0.02) \,a$
(see Fig.\ \ref{fluctVDW}). A relatively large uncertainty is attached
to this value because $\langle \bu(0)^2 \rangle$ varies rapidly in
this regime. Now, we would expect the two techniques to yield
comparable results when choosing $\eta\approx 10\pm 2$. Indeed, the
data compare favorably well: $V^{\eff}_{\eta}(a)$ as determined
from the functional RG with $\eta=8$ exhibits a minimum at
$a/\lambda= 26\pm 2$ (see Fig.\ \ref{veta_frg4}), which is
consistent with the value determined by the variational
procedure. Modified in this way, both techniques hence predict the
same magnetic flux $B_v$ at the transition from the Meissner phase to
the mixed phase.

\section{Conclusion}
\label{sec:concl}

In layered high-$T_c$ superconductors close to the lower melting line
$H_{c_1}$ where the mean flux line distance becomes larger than the
London penetration depth $\lambda$, a long ranged, attractive
interaction between the vortices is induced both by thermal or by
quenched-in disorder fluctuations which has interesting consequences
for the phase diagram of these materials in the regime of low magnetic
fields $B$.  In particular, this van der Waals attraction may lead to a
first order transition from the Meissner phase to the mixed phase,
both for pure and for disordered superconductors.

Restricting ourselves to the pure case, we have established a
variational technique that allows for the computation of the van der
Waals magnetic induction $B_v$, which is based on a self-consistent 
expansion around the regular Abrikosov lattice.  We have applied this
method to the calculation of $B_v$ in the case of BiSCCO,
which is a typical example of a strongly layered high-$T_c$
superconductor. The results for the low field phase diagram 
of pure layered superconductors are qualitatively and, to some extent
also quantitatively, consistent with the results from a functional RG
approach. The latter is the more powerful tool in that it correctly
reproduces logarithmic corrections to the leading scaling behavior.
It is hence adequate for determining the true asymptotic behavior at
the phase transition. The variational technique, on the other hand,
has the advantage that it provides an inherent control mechanism as to
where to stop the coarse graining process or -- in other words --
which is the scale on which the effective interaction has to be
considered. Furthermore, by self-consistently determining the flux
line fluctuations, the results from the variational method can be used
to check {\em a posteriori}, using a Lindemann criterion, whether an
expansion around the regular Abrikosov lattice is justified for a given
set of physical parameters -- which is the basic assumption of the
variational calculation.  Indeed, this has to be checked carefully
since we are working in a part of the phase diagram where the lower
lattice melting line $B_m$ and the van der Waals magnetization $B_v$
are of the same order.

The variational approach may straightforwardly be extended to the
situation with quenched-in impurities using the replica trick. An
advance in this direction could be guided by variational studies of
flux lines lattices in high-$T_c$ superconductors in the presence of
weak disorder \cite{Mezard91,Giamarchi94}.

\begin{acknowledgement}
We thank Jan Kierfeld and Thorsten Emig for helpful discussions and
the German Israeli Foundation (GIF) for financial support. 
\end{acknowledgement}

\end{document}